\documentclass[
prd,
superscriptaddress,
longbibliography,
nofootinbib,
amsmath,
amssymb
]{revtex4-2}

\usepackage{mathtools}
\usepackage{bm}
\usepackage[normalem]{ulem}

\usepackage{graphicx}
\usepackage{xcolor}
\usepackage{booktabs}
\usepackage{array}

\usepackage{tikz}
\usetikzlibrary{arrows.meta,calc,positioning}

\usepackage[
colorlinks=true,
linkcolor=blue,
citecolor=blue,
urlcolor=blue
]{hyperref}

\usepackage[nameinlink,noabbrev]{cleveref}

\begin{document}
\title{Smooth cutoffs and analytic continuation in Casimir physics}

\author{
Lucía N. Helou}
\affiliation{
Instituto Balseiro,  
Centro Atómico Bariloche, R8402AGP Bariloche, Argentina.\\
Instituto de Física Teórica, Universidade Estadual Paulista,
São Paulo, Brazil}
\author{Francisco D. Mazzitelli}
\affiliation{Instituto Balseiro, Centro Atómico Bariloche,
R8402AGP Bariloche, Argentina
}

\begin{abstract}
Divergent series arising in the computation of Casimir energies admit two seemingly different treatments: a physically motivated regularization and subtraction procedure, and formal analytic continuation methods that assign finite values directly to divergent expressions. The agreement between these approaches for the Casimir energy between parallel plates is well known, but its origin is often left implicit. In this work, we provide a unified framework that makes this equivalence transparent. Building on Tao's theory of smoothed sums, we show that the introduction of a smooth cutoff leads to an asymptotic expansion whose finite, regulator-independent part is universally determined by the analytic continuation of the associated Dirichlet series. This identification follows from a Mellin-transform representation, in which divergences and finite contributions are encoded in the pole structure of the integrand. We extend this approach to a class of series relevant to Casimir problems.
In this setting, we show that logarithmic divergences arise from pole coincidences, and we obtain a natural decomposition of the result into universal and cutoff-dependent terms, mirroring the structure of renormalization in quantum field theory.
%These results lead to a Casimir-inspired summability prescription, in which the finite, cutoff-independent part of a regulated series is singled out as its natural value, thereby providing a direct bridge between physical regularization schemes and Ramanujan summation.
\end{abstract}

\maketitle


\section{Introduction}
Divergences are ubiquitous in quantum field theory. Even in the simplest settings, such as the vacuum energy of free fields, formal expressions lead to divergent sums or integrals that require regularization and subtraction in order to extract finite, physically meaningful quantities. A paradigmatic example is provided by the Casimir effect \cite{Milonnibook,Miltonbook,Bordagbook}, where the zero-point energy of a quantum field in the presence of boundaries is formally divergent, yet yields a finite and experimentally verified force after appropriate renormalization.

At the same time, it is well known that formal analytic continuation methods—most notably zeta-function regularization \cite{ElizaldeBook,KirstenBook}—often assign finite values directly to the original divergent expressions. For instance, the identity
\begin{equation}
\sum_{n=1}^\infty n = -\frac{1}{12}
\end{equation}
reproduces the correct Casimir energy in one spatial dimension. This mathematical agreement is routinely used in practice, but its conceptual origin is rarely made explicit. The fact that physical observables (like the force between conductors) are independent of the specific choice of regularizing cutoff is already a well-established feature; therefore, the question we wish to address here is precisely why the surviving finite part  coincides with the value prescribed by analytic continuation.

From a mathematical perspective, this question is closely related to the problem of summability of divergent series. Various methods, including those of Abel, Borel, Cesàro, and Ramanujan, have been developed to associate finite values to divergent expressions (see \cite{Hardy,Cande,overview}). More recently, Tao \cite{Tao} introduced a framework based on smoothed sums, in which a cutoff function is used to define a regulated series, and a finite value is extracted from its asymptotic expansion in a way that is insensitive to the detailed form of the cutoff. Moreover, for the class of series considered in Ref.\cite{Tao}, the finite part coincides with the analytic continuation of the zeta function.

In this work, we show that Tao’s construction provides a natural bridge between the mathematical theory of summability and the physical procedure used in the computation of Casimir energies between parallel plates.
Previous studies have explored connections between smoothed summation and quantum field theory\cite{Wres,Padilla}, as well as the link between exponential cutoffs and zeta regularization via Mellin transforms \cite{BS96}. Here, we extend this correspondence: by employing general smooth cutoff functions within Tao's framework, we show that the regulated sum admits a Mellin-transform representation in which the asymptotic behavior is controlled by the analytic structure of the associated Dirichlet series. In this representation, divergences arise from poles with positive real part, while the finite, cutoff-independent contribution is encoded in the residue at the origin.

This framework allows us to make precise the relation between physical regularization and analytic continuation. For a class of series relevant to Casimir problems, we show that the finite part obtained after regularization and subtraction coincides with the value defined by analytic continuation of the associated Dirichlet series. Moreover, we find that logarithmic divergences emerge from the coincidence of poles in the Mellin representation, leading to a natural separation between universal and cutoff-dependent contributions. This structure closely parallels the organization of divergences in quantum field theory.

We develop these ideas in several steps. We begin by revisiting the standard Casimir effect in one spatial dimension, emphasizing the role of smooth cutoffs and the extraction of finite terms. We then introduce Tao’s smoothed summation method and extend it to a class of series relevant for massive fields and higher dimensions. Using a Mellin-transform representation, we analyze the asymptotic behavior of generic series and classify their divergences in terms of the pole structure of the associated Dirichlet functions.

Motivated by this analysis, one may associate to certain divergent series the finite, cutoff-independent part of their regulated form. This provides a natural link between physical regularization procedures and mathematical summability methods, while clarifying why different prescriptions often lead to the same finite result.

\section{Casimir energy and divergent series}
 When computing the zero point energy $E_0$ for a scalar field satisfying Dirichlet boundary conditions for two
mirrors separated by a distance $a$ in $1+1$ dimensions, one is formally led to the divergent series
\begin{equation}
E_0=\frac {\pi}{ 2 a}\sum_{n=1}^\infty\ n\, .
\end{equation}
This is the contribution to the vacuum energy coming from the fluctuations of the quantum field between the perfect mirrors. The definition of the zeta function $\zeta(s)$
is, for $s>1$, 
\begin{equation}
\zeta(s)=\sum_{n=1}^\infty\ n^{-s}\, .
\end{equation}
The analytic continuation gives $\zeta(-1) = -\frac{1}{12}$, 
so one gets for the Casimir energy $E_C$
\begin{equation}
E_c(a)=\frac {\pi}{ 2 a}\zeta(-1)=-\frac{\pi}{24 a}\, .
\end{equation}

Although this is the correct result, the derivation is not well justified from a physical point of view. As reviewed in \cite{Greiner}, the steps to obtain a finite answer for the Casimir force involve both a regularization and a subtraction. On the one hand, for a realistic material, the perfect conductor boundary conditions do not apply to the ultraviolet modes (the mirrors become transparent for high frequency modes).  This physical property can be taken into account
by introducing a smooth cutoff in the series. 
On the other hand, one must also consider the contribution of the zero-point energy coming from the outside region. Finally, only differences of energies are measurable (the force is the derivative of $E_C$ with respect to $a$). Therefore,  $a$-independent divergences are irrelevant and the force is independent of the cutoff.

We describe here the calculation. We assume the presence of mirrors at $x=0,\,  x=a$ and $x=L$ with $L\gg a$ and impose Dirichlet boundary conditions on them. Assuming an exponential cutoff, the vacuum energy is the limit $N\to \infty$ of
\begin{equation}
E_0(a)= \frac {\pi}{ 2 a}\sum_{n=1}^\infty\ n e^{-\frac{n\lambda_0}{N a}}+\frac {\pi}{ 2 (L-a)}\sum_{n=1}^\infty\ n e^{-\frac{n\lambda_0}{N(L-a)}}\, .
\end{equation}
 Here $\omega_c =N\pi/(2\lambda_0)$  denotes the cutoff frequency, with $\lambda_0$ a constant with units of length.

The Casimir energy $E_C(a)$ is the difference between the vacuum energy for the mirrors separated at a distance $a$ and that corresponding to a distance much larger than $a$, that is
\begin{equation}\label{E_C}
    E_C(a)=E_0(a)-E_0(L/d)\, ,
\end{equation}
where $d$ is $O(1)$ and greater than $1$, in such a way that $L/d$ is still much larger than $a$.

An explicit calculation gives
\begin{equation}\label{reg-sum}
\sum_{n=1}^\infty\ n e^{- \alpha n/a}=\frac{a^2} {\alpha^2}-\frac{1}{12 }+O\left(\alpha^2\right)
\end{equation}
and therefore
\begin{equation}
    E_C(a)=-\frac{\pi}{24 a} + O(a/L)\, .
\end{equation}
Note that the divergent term in $E_0(a)$ is proportional to $L$, and this is the mathematical reason  of the cancellation of the divergences in Eq.\eqref{E_C}.

The crucial point for the agreement between the analytic continuation and the regularized computation of the Casimir energy is that, in the regularized sum (\ref{reg-sum}), the $\alpha$- independent term is the analytic continuation 
$\zeta(-1)=-1/12$. This is not merely a numerical coincidence, but a general property of the regularized series. Moreover, the result is independent of the cutoff function, as long as it is sufficiently smooth. 

%Similar arguments give the well known result for the Casimir effect in $3+1$ dimensions for the electromagnetic field in the presence of two planar perfect conductors. 

\section{Smoothed series}

Tao's main idea is to replace the abruptly truncated partial sums by smoothed sums
\begin{equation}
    \sum_{n=1}^Na_n \rightarrow \sum_{n=1}^\infty a_n\eta(n/N)
\end{equation}
where $\eta(t)$ is a cutoff function that satisfies $\eta(0)=1$ and $\eta(1)=0$, i.e.,  it has a compact support. The smoothing does not affect the value of series that are absolutely convergent.

When applying the smoothing to otherwise divergent series one can show that [4]
\begin{eqnarray}
    \sum_{n=1}^\infty \eta(n/N) &=& -\frac{1}{2}+C_0 N + O(1/N)\nonumber\\
    \sum_{n=1}^\infty n \eta(n/N) &=& -\frac{1}{12}+C_1 N^2 + O(1/N)\nonumber\\   
    \sum_{n=1}^\infty n^2\eta(n/N) &=& 0 + C_2 N^3 + O(1/N) \label{smoothed}   
    \end{eqnarray}
where the $C_i$ and the terms $O(1/N)$ depend on the cutoff function, but the constant terms, independent of $N$, do not. Recalling that $\zeta(0)=-1/2, \zeta(-1)=-1/12$ and $\zeta(-2)=0$, we see that the results obtained by analytic continuation of the $\zeta$ function are nothing more than the constant terms of the asymptotic expansion of the smoothed partial sums (note that this procedure does not work for a sharp cutoff function). 

In order to prove the results in Eq.\eqref{smoothed}, we  relax the compact support hypothesis for the cutoff function, and assume that it is a Schwartz function, i.e. 
 a $C^\infty$ function $\eta(t)$ that tends to zero as $\vert t\vert\to\infty$ faster than any inverse power of $t$.
Writing the cutoff function in terms of its Mellin transform $C_\eta(s)$:
\begin{equation}
\eta(t)=\frac{1}{2\pi i}\int_{c-i\infty}^{c+i\infty} ds\, t^{-s}\, C_\eta(s)\, .
\end{equation}
 we obtain
\begin{eqnarray}\label{smoothMellin}
\sum_{n=1}^\infty n^\beta\eta(n/N)&=&\frac{1}{2\pi i}\sum_{n=1}^\infty \int_{c-i\infty}^{c+i\infty} ds\, N^s n^{\beta-s}\, C_\eta(s)\nonumber\\
&=& \frac{1}{2\pi i}\int_{c-i\infty}^{c+i\infty} ds\, N^s\zeta(s-\beta)C_\eta(s)\, .
\end{eqnarray}

If $\eta$ is a Schwartz function, the Mellin transform  
$C_\eta(s)$ admits a meromorphic extension  that has simple poles 
at $s_{n}=-n$ ($n=0,1,...$). This is shown in the Appendix. Moreover, if  $\eta(t) = \sum_{n=0}^{\infty} \eta_n t^n$, the residue of $C_\eta(s)$ at $s_{n}$ 
is $\eta_n$ (we recall that $\eta_0=1$). On the other hand, the $\zeta(z)$-function has a single pole at $z=1$.  Therefore the integral in Eq.\eqref
{smoothMellin} can be evaluated using tne Cauchy theorem. We choose $c$ to be at the right of all singularities of the integrand and close the integration contour to the left ($\text{Re}(s) \to -\infty$). Since the integrand contains the factor $N^{s}$, this contour shift is mathematically justified, and the integral is determined by the sum of the residues of the poles of $C_\eta(s) \zeta(s-\beta)$ in the complex plane.
We obtain
\begin{equation}\label{proof}
\sum_{n=1}^\infty n^\beta\eta(n/N)=
N^{1+\beta}C_\eta(1+\beta)+\zeta(-\beta)+O(1/N)\, .
\end{equation}
This proves Eq.\eqref{smoothed}.

Tao's arguments apply to the Casimir physics. To see this, the series we shall consider is the following:
\begin{equation}
    \sum_{n=0}^\infty \omega_n \eta \left( \frac{\omega_n}{\omega_c} \right),
\end{equation}
where $\omega_n=\pi n/(2a)$ and $\omega_c$ is the cutoff frequency introduced before. Note that
we are assuming that the cutoff function depends on the whole coefficient of the series. In this particular case this fact is trivial, although it will not be trivial for more general series. The proposal is of course based on physical considerations, and will be adequate or not depending on the properties of the materials. Thus we have
\begin{equation}
     E(a)=\frac{\pi}{2a}\sum_{n=0}^\infty n \, \eta \left( \frac{\lambda_0}{a}\frac{n}{N} \right).
\end{equation}
Using Eq. \eqref{proof} we obtain
\begin{equation}
    E(a)=\frac{\pi}{2 a}\left[ \left(\frac{aN}{\lambda_0}\right)^2 C_\eta(2) -\frac{1}{12}+O\left(\frac{1}{N}\right)\right].
\end{equation}
As $E_0(a)=E(a)+E(L-a)$,  the divergent (and cutoff-dependent) term in $E_0(a)$ depends linearly on $L$, and therefore will cancel out after the subtraction (see Eq.\eqref{E_C}), giving the usual result for the Casimir energy in the limit $N\to\infty$.

The fact that the finite part of the series coincides with the analytic continuation $\zeta(-1)$ can be illustrated explicitly for different choices of smooth cutoff functions. For instance, one finds
\begin{equation}
    \frac{\pi}{2a}\sum_{n=0}^\infty n e^{-n\lambda_0/aN}=\frac{\pi}{2 a}\left[\left(\frac{aN}{\lambda_0}\right)^2 -\frac{1}{12}+O\left(\frac{1}{N}\right)\right],
\end{equation}

\begin{equation}
    \frac{\pi}{2a}\sum_{n=0}^\infty \frac{n}{(1+(n\lambda_0/aN))^3}=\frac{\pi}{2 a}\left[\frac{1}{2}\left(\frac{aN}{\lambda_0}\right)^2-\frac{1}{12}+O\left(\frac{1}{N}\right)\right],
\end{equation}

\begin{equation}
    \frac{\pi}{2a}\sum_{n=0}^\infty n\left(1-\frac{\lambda_0 n}{aN}\right)^2 \Theta\left(1-\frac{\lambda_0 n}{aN}\right) = \frac{\pi}{2 a}\left[\frac{1}{12} \left(\frac{aN}{\lambda_0}\right)^2 -\frac{1}{12}\right].
\end{equation}
In all cases, the finite, cutoff-independent part is the same, and coincides with the result obtained with analytic continuation. This property is valid when the cutoff function and its first derivative are continuous. By contrast,
if the derivative is discontinuous, the constant term may change.  For example 
\begin{equation}
    \frac{\pi}{2a}\sum_{n=0}^\infty n\left(1-\frac{\lambda_0 n}{aN}\right) \Theta\left(1-\frac{\lambda_0 n}{aN}\right) = \frac{\pi}{2 a }\left[\frac{1}{6}\left(\frac{aN}{\lambda_0}\right)^2  -\frac{1}{6 }\right].
\end{equation}
 
 These examples show that the analytic continuation yields the same result as the smoothing procedure, as long as the smooth cutoff offers an appropriate representation of the underlying physics.
 The general origin of this universality can be understood in terms of a Mellin transform representation of the cutoff function, in which the asymptotic behavior is governed by the singularities of the corresponding Dirichlet series. We generalize this approach in the next sections.

\section{A class of generalized smoothed series}
We now discuss a generalization of Tao's smoothed series approach.
The Casimir energy for a massless scalar field in $3+1$ dimensions is formally given by the series 
\begin{equation}
    \frac{\pi}{2 a}\sum_{n,l,p=1}^\infty \left(n^2+\frac{l^2}{L_1^2}+\frac{p^2}{L_2^2}\right)^{1/2}=\frac{\pi}{2a}\sum_{l,p=1}^\infty \left( \sum_{n=1}^\infty (n^2+b^2_{l,p})^{1/2}\right),
\end{equation}
while for a massive field of mass $m$ in $1+1$ dimensions we have a series of the form
\begin{equation}\label{campo con masa primero}
    \frac{\pi}{2a}\sum_{n=1}^\infty \left(n^2+\left(\frac{ma}{\pi}\right)^2\right)^{1/2}.
\end{equation}

From these examples we see that, in order to make a connection with more realistic calculations of the Casimir energy, the smoothed series approach to summability should be generalized. We will consider series of the form
\begin{equation}\label{general S(alpha,beta)}    S(\beta,b)=\sum_{n=1}^\infty \left(n^2+b\right)^{\beta}\eta\left(\frac{n^2+b}{N^2}\right),
\end{equation}
where $\beta$ and $b$ are real constants, with  $b>0$.  
As we will see, different values of $\beta$ change drastically the qualitative behavior of the series. 

Using the Mellin inverse transform
\begin{equation}\label{antitransformada}
    \eta \left( \frac{n^{2} + b}{N^{2}} \right) = \frac{1}{2 \pi i} \int_{c - i \infty}^{c + i \infty} ds\,  C_\eta(s) \left( \frac{n^{2} + b}{N^{2}} \right)^{-s}.
\end{equation}
we rewrite the series as
\begin{eqnarray}\label{INTEGRAL RESIDUOS}
   S(\beta,b)&=& \frac{1}{2 \pi i} \int_{c - i \infty}^{c + i \infty} ds\,  N^{2s} C_\eta(s)\sum_{n=1}^\infty \left( n^{2} + b \right)^{\beta-s} \nonumber\\
   &=&\frac{1}{2 \pi i} \int_{c - i \infty}^{c + i \infty} ds\,  N^{2s} C_\eta(s) Z(s-\beta,b),  
\end{eqnarray}
where $Z(z,b)$ is the one-dimensional inhomogeneous Epstein-zeta function \footnote{Other  notations for this Epstein-zeta function  are $E^{b}_1(s)$ or $E_1(s,b)$, see for instance \cite{Kirsten91,Elizalde94}.}
\begin{equation}
Z(z,b) = \sum_{n=1}^\infty (n^2+b)^{-z},
\end{equation}
that has simple poles at $z = 1/2, -1/2, -3/2, \dots$.  

The integral in Eq.\eqref{INTEGRAL RESIDUOS} can be evaluated using the the same procedure as before (see the discussion after Eq.\eqref{smoothMellin}).
It is determined by the sum of the residues of the poles of $C_\eta(s) Z(s-\beta, b)$ in the complex plane:
\begin{equation}
S(\beta,b) = \sum_{s_k} \text{Res}\left[ C_\eta(s) Z(s-\beta, b) N^{2s}, s_k \right]\, .
\end{equation}
The location of these poles determines the physical nature of each term in the asymptotic expansion. If all singularities are simple poles, 
those located at $\text{Re}(s) > 0$ yield terms proportional to positive powers of $N$, while poles located at $\text{Re}(s) < 0$ produce negative powers of $N$, which  vanish in the limit $N \to \infty$.

As the Mellin transform $C_\eta(s)$ possesses a simple pole at $s=0$ with residue $1$,  if the  function $Z(s-\beta, b)$ is analytic at $s=0$, the finite part of the summation is completely independent of the shape of the regulator and is given  by its analytic continuation, $Z(-\beta, b)$.
This establishes the equivalence between the physical smooth cutoff regularization and the zeta function regularization for this class of series.

However, a mathematically and physically richer scenario occurs if $Z(s-\beta, b)$ also exhibits a simple pole at $s=0$ (which happens when 
$\beta = -\tfrac{1}{2},\allowbreak\, \tfrac{1}{2},\allowbreak\, \tfrac{3}{2},\allowbreak\, \ldots$). In this case, the coincidence of both poles creates a double pole at $s=0$ for the full integrand. 
We can calculate this contribution explicitly by performing a Laurent expansion of the integrand around $s=0$. For the cutoff function, we have:
\begin{equation}
C_\eta(s) = \frac{1}{s} + c_\eta + \mathcal{O}(s)\, ,
\end{equation}
where $c_\eta$ is a regulator-dependent constant. The Epstein-zeta function has a simple pole at $s=0$ with residue $R_0(\beta)$ given by \cite{Toms}
\begin{equation}
R_0(\beta)=\frac{1}{2}\left(\frac{b}{4}\right)^{\beta+1/2}\frac{(2\beta+1)!}{(\beta+1/2)!^2},
\end{equation}
and therefore
\begin{equation}
Z(s-\beta, b) = \frac{R_0(\beta)}{s} + Z_{reg}(-\beta, b) + \mathcal{O}(s)\, ,
\end{equation}
where $Z_{reg}(-\beta, b)$ is the regular, analytic part of the Epstein-zeta function evaluated at $s=0$. Finally, the $N^{2s}$ factor expands as:
\begin{equation}
N^{2s} = e^{2s \ln N} = 1 + 2s \ln N + \mathcal{O}(s^2)\, .
\end{equation}
Multiplying these three series together, the integrand around $s=0$ has a double pole $\mathcal{O}(1/s^2)$ and a simple pole $\mathcal{O}(1/s)$. By the residue theorem, the integral's evaluation at this pole is given  by:
\begin{equation}\label{eq:double_pole_residue}
\text{Res}\left[ C_\eta(s) Z(s-\beta, b) N^{2s}, s=0 \right] = 2 R_0(\beta) \ln N + Z_{reg}(-\beta, b) + c_\eta R_0(\beta)\, .
\end{equation}
 Note that the universal finite part (i.e. the finite part independent of the cutoff) can be written as 
\begin{equation}
Z_{reg}(-\beta, b) = \frac{d}{ds}\left(s Z(s-\beta,b)\right)\vert_{s=0}\, .
\end{equation}
Eq.\eqref{eq:double_pole_residue}  isolates three distinct physical contributions: a cutoff-independent logarithmic divergence, a universal finite part related to the analytic continuation of the Epstein zeta-function, and a finite term
that depends explicitly  on the specific choice of the smooth cutoff profile.

\subsection{The Massive Casimir Effect in 1+1 Dimensions}

To illustrate the mechanism of pole coincidence and trace the dependence on the physical scales, we analyze the zero-point energy of a massive scalar field of mass $m$, confined in a 1D box of length $a$. For Dirichlet boundary conditions, the discrete frequencies are given by $\omega_n = \sqrt{k_n^2 + m^2}/2$, where $k_n = n\pi/a$. Introducing a smooth, high-energy cutoff function $\eta(\omega/\omega_c)$ (where, as before, $\omega_c=N\pi/(2\lambda_0)$ is the cutoff scale), the regularized vacuum energy reads \footnote{ A similar calculation for the particular case of an exponential cutoff and periodic boundary conditions can be found in Ref.\cite{Teo}}:
\begin{equation}
E(a) =  \sum_{n=1}^{\infty} \omega_n \, \eta\left(\frac{\omega_n}{\omega_c}\right) = \frac{\pi}{2a} \sum_{n=1}^{\infty} \sqrt{n^2 + b} \, \eta\left( \frac{\lambda_0}{a N} \sqrt{n^2 + b} \right)\, ,
\end{equation}
where we have factored out $\pi/a$ and defined the dimensionless mass parameter $b = (ma/\pi)^2$. 

Using the Mellin representation of $\eta(t)$, the regularized energy takes the exact form of our previous integral:
\begin{equation}
E(a) = \frac{\pi}{2a} \left[ \frac{1}{2\pi i} \int_{c-i\infty}^{c+i\infty} ds \, C_\eta(s) Z(s-1/2, b) \left(\frac{a N}{\lambda_0}\right)^{2s} \right]\, .
\end{equation}

Closing the contour to the left for $N \to \infty$, we encounter the poles. The first is a simple pole at $s=1$ originating from $Z(s-1/2,b)$, which dictates the primary bulk volume divergence:
\begin{equation}
E_{s=1} = \frac{\pi}{2a} C_\eta(1) \text{Res}[Z(s-1/2,b), 1] \left(\frac{a N}{\lambda_0}\right)^2 = \left[ \frac {\pi}{4}C_\eta(1)\left(\frac{N}{\lambda_0} \right)^2 \right] a \, .
\end{equation}
As expected, this leading divergence scales linearly with the volume $a$.

The most subtle physics lies at $s=0$. Here, the cutoff function $C_\eta(s)$ has a simple pole, but the Epstein-zeta function $Z(s-1/2,b)$ also possesses a simple pole at $s=0$ with residue  $R_0(1/2) = b/4 = \frac{m^2 a^2}{4\pi^2}$.
Multiplying the residue of the resulting double pole derived in the previous section by the overall factor $\pi/2a$, we find the $s=0$ contribution:
\begin{equation}
E_{s=0} = \frac{\pi}{2a} \left[ \frac{b}{2} \ln (aN/\lambda_0) + \frac{b}{4}c_\eta + Z_{reg}(-1/2, b) \right]\, .
\end{equation}
To compute $Z_{reg}(-1/2, b)$ explicitly, one employs the Poisson summation formula \cite{Elizalde2000} to analytically continue the series $\sum (n^2+b)^{-s}$. Evaluating the regular part at $s=-1/2$ yields:
\begin{equation}
Z_{reg}(-1/2, b) = -\frac{b}{4} \left( 1 + \ln\frac{b}{4} \right) - \frac{\sqrt{b}}{2} - \frac{\sqrt{b}}{\pi} \sum_{k=1}^\infty \frac{1}{k} K_1(2\pi k \sqrt{b})\, ,
\end{equation}
where $K_1(z)$ is the modified Bessel function of the second kind. Recalling that $b = (ma/\pi)^2$,  the full contribution from the $s=0$ pole becomes:
\begin{align}
E_{s=0} = \frac{\pi}{2a} \Bigg[ & \frac{m^2 a^2}{2\pi^2} \ln \left(\frac{aN}{\lambda_0}\right) + c_\eta \frac{m^2 a^2}{4\pi^2} \nonumber \\
& - \frac{m^2 a^2}{4\pi^2}\left(1 + \ln \frac{m^2 a^2}{4\pi^2}\right) - \frac{ma}{2\pi} - \frac{ma}{\pi^2} \sum_{k=1}^\infty \frac{1}{k} K_1(2mak) \Bigg]\, .
\end{align}

A remarkable cancellation occurs between the logarithmic divergence and the analytic continuation of the zeta function. Combining the logarithms within the bracket, the non-linear $a \ln a$ dependence exactly vanishes, leaving:
\begin{equation}\label{eq:Es=0}
E_{s=0} = \frac{m^2 a}{8\pi} \left( 2\ln \frac{2 \pi N}{\lambda_0m} - 1 + c_\eta \right) - \frac{m}{4}- \frac{m}{2\pi} \sum_{k=1}^\infty \frac{1}{k} K_1(2 m a k) .
\end{equation}
It is worth  remarking that, for the particular case of an exponential regulator $\eta(t)=\exp[-t]$,  one has $c_\eta = -\gamma$. In this case,  Eq.\eqref{eq:Es=0} accounts for the $-\gamma$ difference  in the finite part between the exponential cutoff and zeta-function regularization found in Ref.\cite{BS96}.

This explicit evaluation  exposes the physics of the regularized massive vacuum energy. The pole coincidence at $s=0$ indeed produces a logarithmic divergence and a cutoff ambiguity $c_\eta$, but these renormalize the extensive bulk energy (linear in $a$). The finite term $-m/4$ represents a constant energy that is not relevant for computing forces. Consequently, the Casimir force is entirely governed by the rapidly decaying Bessel functions \cite{alvarez09}, remaining perfectly universal and isolated from the cutoff ambiguity.

\section{Classification of divergent series}

In  Tao's original formulation, for series in which $a_n$ is a power of $n$, smooth summation leads to a power-law divergence with a cutoff-dependent coefficient, together with a cutoff-independent finite part. For the class of series considered in the previous section, the structure can be richer: besides power divergences, cutoff-independent logarithmic divergences may arise, and the finite part generally splits into a universal and a cutoff-dependent contribution. In the following we analyze this behavior in greater generality.

Consider the series 
\begin{equation}
    S_\eta(a_c)=\sum_{n=1}^\infty a_n\eta\left (\frac{a_n}{a_c}\right),
\end{equation}
where $a_c$ is a large cutoff parameter, \(\eta(t)\) is a Schwartz function satisfying
$\eta(0)=1 $,  and  the sequence ${a_n}$ is positive definite and monotonically increasing.
We already mentioned that the Mellin transform $C_\eta(s)$
is well defined for  \(\Re(s)>0\), and admits a meromorphic continuation to the complex plane with simple poles at $s=-n$.
We also assume that the  generalized Dirichlet series
\begin{equation}
D_g(s)=\sum_{n=1}^\infty a_n^{1-s}
\end{equation}
admits a meromorphic continuation to the complex plane. 
 
 Using the Mellin representation of the cutoff, the regulated sum can be written as
\begin{equation}
S_\eta(a_c)=\frac{1}{2\pi i}\int_{c-i\infty}^{c+i\infty} ds\, a_c^s\, C_\eta(s)\, D_g(s),
\end{equation}
where, as in the previous section, \(c\) is chosen to the right of all singularities of the integrand. 
The divergence structure of the series $S_\eta(a_c)$ is entirely determined by the singularities of \(D_g(s)\) and $C_\eta(s)$ (note that the last function contributes a simple pole at \(s=0\) with unit residue). The analysis of the behavior of the series is the following:

\begin{itemize}
\item Case 1: No poles of \(D_g(s)\) for \(\Re(s)\ge0\). \\
The only pole of the integrand is that of $C_\eta(s)$ at $s=0$.  Therefore the regulated sum is $O(a_c^0)$ in the limit \(a_c\to\infty\). The series is convergent and the result is independent of the cutoff.

\item Case 2: Simple pole of \(D_g(s)\) at \(s=\alpha>0\). \\
The regulated sum diverges as \(a_c^\alpha\). The coefficient of the power divergence depends on the cutoff shape, while the finite term, determined by the pole at $s=0$, does not depend on \(\eta\) and it is given by the analytic continuation $D_g(0)$. The particular case in which $a_n = n^p, \, p=0,1,2,\cdots$ was considered in Ref.\cite{Tao}.

\item Case 3: Simple pole of \(D_g(s)\) at \(s=0\). \\
The product \(C_\eta(s)D_g(s)\) develops a double pole at the origin, leading to a logarithmic divergence \(\sim \log a_c\) with a coefficient that is independent of the cutoff.  The (cutoff-independent) finite part is given by
\begin{equation}
\frac{d}{ds}(sD_g(s))\vert_{s=0}
\end{equation}
This is the case for the series described in the previous section. 

\item Case 4: Higher-order poles. \\
If \(D_g(s)\) has a pole of order \(k\) at \(s=\alpha\), the regulated sum contains terms of the form
\begin{equation}
a_c^\alpha (\log a_c)^{k-1},
\end{equation}
The coefficients are determined by the singular terms in the Laurent expansion of $D_g(s)$ around the pole and depend on the cutoff when $\alpha\neq 0$. Poles of higher order at \(s=0\) generate higher powers of logarithmic divergences.
\end{itemize}

This classification parallels the familiar distinction between power and logarithmic ultraviolet divergences in quantum field theory. While previous formulations based on heat kernel expansions (such as Ref.\cite{BS96}) frame the emergence of logarithmic divergences in terms of specific heat kernel/Seeley coefficients on Riemannian manifolds, the classification scheme presented here applies more broadly to general spectrum sequences $a_n$ and arbitrary smooth cutoff profiles $\eta(t)$. By focusing directly on the pole structure of the generalized Dirichlet series $D_g(s)$, this approach encompasses not only boundary-value problems governed by elliptic differential operators, but also wider classes of divergent series in quantum field theory and number theory where heat kernel techniques may not be directly applicable.

From the mathematical side, closely related results appear in Tauberian theory (see e.g. Ref.\cite{Korevaar} and the recent discussion in\cite{Pierce}), where the analytic structure of Dirichlet series controls the asymptotic behavior of partial sums. In this context, the presence and multiplicity of poles determine the emergence of power-law and logarithmic contributions in asymptotic expansions. The present approach differs from the traditional Tauberian setting in that it makes systematic use of smooth cutoffs and emphasizes the extraction of finite parts beyond the leading asymptotic behavior, allowing for a clear separation between universal and cutoff-dependent contributions. There is also 
a heuristic discussion of the relation between the poles of the Dirichlet function and the asymptotic behavior of the series  in Ref.\cite{Tao}, where  only the Case 2 is briefly mentioned. We discuss now some examples.

\subsection{The harmonic series \texorpdfstring{$\sum 1/n$}{harmonic series}}
Consider 
\[
S_\eta(N)=\sum_{n=1}^\infty \frac{1}{n}\,\eta\!\left(\frac{n}{N}\right).
\]
Using the inverse Mellin representation,
\[
\eta\!\left(\frac{n}{N}\right)
=\frac{1}{2\pi i}\int_{\Re(s)=c>0} C_\eta(s)\,N^s\,n^{-s}\,ds,
\]
we obtain
\[
S_\eta(N)=\frac{1}{2\pi i}\int_{\Re(s)=c}
C_\eta(s)\,\zeta(1+s)\,N^s\,ds.
\]

The singular structure near $s=0$ is given by
\[
\zeta(1+s)=\frac{1}{s}+\gamma+O(s), \qquad
C_\eta(s)=\frac{1}{s}+c_\eta+O(s),
\]
together with
\[
N^s=1+s\log N+\frac{s^2}{2}(\log N)^2+\cdots .
\]
Their product yields
\[
C_\eta(s)\,\zeta(1+s)\,N^s
=
\frac{1}{s^2}
+\frac{\log N+\gamma+c_\eta}{s}
+O(1),
\]
so that the integrand has a double pole at $s=0$.
Evaluating the contour integral by residues, we find
\[
S_\eta(N)=\log N+\gamma+c_\eta+o(1), \qquad N\to\infty.
\]

The logarithmic divergence is universal and independent of the cutoff, while the constant term naturally splits into a universal part, given by the Euler--Mascheroni constant $\gamma$, and a non-universal contribution $c_\eta$ depending on the detailed form of the regulator.
Subtracting both the divergent term and the cutoff-dependent finite constant, one is left with the universal finite part (UFP)
\[
\text{UFP}\left(\sum_{n=1}^\infty \frac{1}{n}\right) \;=\; \gamma.
\]
Note that for this series zeta-function regularization fails. It is also interesting to remark that 
this finite result coincides with the Euler-MacLaurin constant of the series, also called the Ramanujan sum of the series \cite{Hardy}.

\subsection{The logarithmic series \texorpdfstring{$\sum \log n$}{logarithmic series}}
 The generalized series $D_g(s)$ corresponding to $a_n=\log n$
 does not converge for any value of $s$, and no natural meromorphic continuation is available. As a result, the residue-based analysis breaks down, and the cutoff-dependent terms cannot be disentangled from a universal finite contribution using this type of cutoff. However, we can obtain a finite answer for this series by considering
\begin{equation}
S_\eta(N)=\sum_{n=1}^\infty \log n\, \eta\!\left(\frac{n}{N}\right).
\end{equation}
The associated Dirichlet series is
\begin{equation}
D(s)=\sum_{n=1}^\infty \log n\, n^{-s} = -\zeta'(s),
\end{equation}
which has a double pole at \(s=1\),  with expansion
\begin{equation}
-\zeta'(s)=\frac{1}{(s-1)^2}+\gamma_1+\cdots,
\end{equation}
where $\gamma_1$ is the Stieljes constant.
Since $C_\eta(s)$ is regular at \(s=1\), the integrand \( C_\eta(s)D(s)N^s\) has a double pole at \(s=1\). Evaluating the residue yields
\begin{equation}
\text{Res}\left[ C_\eta(s) \zeta'(s) N^{s}, s=1 \right] =-N\left(C_\eta(1)\log N+C'_\eta(1)\right).
\end{equation}
On the other hand
\begin{equation}
\text{Res}\left[ C_\eta(s) \zeta'(s)  N^{s}, s=0 \right] =\zeta'(0)= -\frac{1}{2}\log(2\pi).
\end{equation}
Therefore the UFP reads
\begin{equation}
\text{UFP}\left(\sum_{n=1}^\infty \log n\right)=\frac{1}2{}\log(2\pi)\, .
\end{equation}
which again coincides with the Ramanujan sum of the series \cite{Hardy}.

The analysis presented in this Section shows that, for a class of divergent series, one can isolate a finite contribution that is independent of the cutoff. In the examples discussed above, this finite part coincides with the values assigned to the same series by other summability prescriptions, such as Ramanujan summation.

There is, however, an important subtlety. In the context of Casimir physics, the cutoff is naturally implemented in terms of the spectral variable. From a physical viewpoint, it should be a function of a
$a_n/a_c$, where $a_c\to\infty$ plays the role of a cutoff scale. For some cases (like the logarithmic series) 
the associated Dirichlet series $D_g(s)$ does not
admit a meromorphic continuation to the complex plane. In this situation, one can still assign a finite value to it, by modifying the structure of the cutoff function. 
%However, the logarithmic example discussed above exhibit a crucial difference. While the standard Dirichlet series associated with $a_n=n^p$ can be analytically continued and has a simple pole structure, the generalized series $D_g(s)$ corresponding to $a_n=\log n$ does not converge for any value of $s$, and no natural meromorphic continuation is available. As a result, the residue-based analysis breaks down, and the cutoff-dependent terms cannot be disentangled from a universal finite contribution using this type of cutoff.

\section{Conclusions}

 In this work we have clarified the relation between the evaluation of divergent series in quantum field theory and Tao’s summability method based on smoothed sums. By introducing a smooth cutoff and analyzing the resulting asymptotic expansion, we have shown that the finite, regulator-independent part of the series coincides with the value obtained by analytic continuation of the associated Dirichlet function.

The key element underlying this correspondence is a Mellin-transform representation of the regulated sum. In this framework, the asymptotic behavior is entirely determined by the pole structure of the product between the Mellin transform of the cutoff and the relevant Dirichlet series. Divergent contributions arise from poles with positive real part, while the finite term is associated with the contribution at the origin. In particular, logarithmic divergences emerge from the coincidence of poles, providing a simple and general interpretation of their origin.

We have extended Tao’s construction to a class of series relevant for computing the Casimir energy between parallel plates, including massive fields and higher dimensions. In this setting, the structure of divergences closely parallels that of renormalization in quantum field theory: power-law divergences depend on the cutoff, logarithmic divergences are universal, and the finite part splits into a universal contribution and a regulator-dependent term.

These results also help place several familiar prescriptions for divergent series within a common framework. In the examples considered here, the finite part independent of the smooth cutoff, analytic continuation, and other summability methods lead to the same finite result.

\section* {Acknowledgments}
This research was supported by  Consejo Nacional de Investigaciones Científicas y Técnicas (CONICET) and Universidad Nacional de Cuyo (UNCuyo).

%%%%%%%%%%%%%%%%%%%%%%%%%
%%%%%%%%%%%%%%%%%%%%%%%%%
\section*{Appendix}

In this Appendix we show that, given  a Schwartz function $\eta(t)$, its Mellin transform $C_\eta(s)$ is a meromorphic function with poles at $s=-n$, where $n=0,1,2,\cdots$ (we remind that a Schwartz function is a $C^\infty$ function $\eta(t)$ that tends to zero as $\vert t\vert\to\infty$ faster than any inverse power of $t$).

We first write the Mellin transform as
\begin{equation}
    C_\eta(s)=\int_0^1 dt \, t^{s-1}\eta(t)+\int_1^\infty dt \, t^{s-1}\eta(t),
\end{equation}
and then rewrite it as
\begin{equation}
\begin{split}
        C_\eta(s)= \int_0^1 & dt \left( \eta(t) - \sum_{n=0}^{M-1} \eta_n t^{\ n} \right) t^{s-1}  + \\
        &+ \sum_{n=0}^{M-1} \frac{\eta_n}{ n + s} + \int_1^{\infty} dt \, \eta(t) t^{s-1} ,
\end{split}\label{C_fsub}
\end{equation}
where the $\eta_n$ are the coefficients in the series expansion of $\eta(t)=\Sigma_{n=0}^{\infty}\eta_nt^n$. As the difference in the first term in Eq.\eqref{C_fsub} is
$O(t^{M})$, the first integral converges for
$\Re(s) > -M $.
On the other hand, taking into account that $\eta(t)$ is a Schwartz function, the second integral converges $\forall s \in \mathbb{C}$ . Taking the limit $M\to\infty$, we see that  \( C_\eta(s) \) has a meromorphic extension in the complex plane, given by 
\begin{equation}
C_\eta(s)=\sum_{n=0}^{\infty} \frac{a_n}{n + s} + \int_1^{\infty} f(t) t^{s-1} \, dt
\end{equation}
with simple poles at  \( s = -n \) (\( n = 0, 1 \ldots\)) with residues $\eta_n$.

\end{document}